\documentclass[11pt]{article}
\begin{document}

\title{{\bf Observational Selection Effects \\ in Quantum Cosmology}
\thanks{Alberta-Thy-23-07, arXiv:0712.2240, to be published in {\it
Proceedings from the 13th International Congress of Logic, Methodology
and Philosophy of Science, Tsinghua University, Beijing, China, August
9-15, 2007}, edited by Clark Glymour, Wei Wang, and Dag
Westerst{\aa}hl (King's College Publications, London).}}

\author{
Don N. Page
\thanks{Internet address:
don@phys.ualberta.ca}
\\
Institute for Theoretical Physics\\
Department of Physics, University of Alberta\\
Room 238 CEB, 11322 -- 89 Avenue\\
Edmonton, Alberta, Canada T6G 2G7
}

\date{(2007 December 13)}

\maketitle
\large

\begin{abstract}
\baselineskip 18 pt

Scientific theories need to be testable by observations, say using
Bayes' theorem.  A complete theory needs at least the three parts of
dynamical laws for specified physical variables, the correct solution
of the dynamical laws (boundary conditions), and the connection with
observations or experience or conscious perceptions (laws of
psycho-physical parallelism).  Principles are proposed for Bayesian
meta-theories.  One framework that obeys these principles is Sensible
Quantum Mechanics (SQM), which is discussed.  In principle, it allows
one to test between single-history and many-worlds theories, and to
discuss threats to certain theories from fake universes and Boltzmann
brains.  The threat of fake universes may be dismissed if one doubts
the substrate-independence of consciousness, which seems very
implausible in the SQM framework.  Boltzmann brains seem more
problematic, though there are many conceivable solutions.  SQM also
suggests the possibility that past steps along our evolutionary
ancestry may be so rare that they have occurred nowhere else within
the part of the universe that we can observe.

\end{abstract}
\normalsize
\baselineskip 15.8 pt
\newpage

\section{Goals and criteria for scientific theories}

I see science as having the following goals:

\begin{enumerate}
\item Theories to explain observations
\item Observation-weighted probabilities for these theories
\item Predictions for these theories
\item Understanding of these theories
\end{enumerate}

A complete theory or model of the universe needs at least three parts:

\begin{enumerate}
\item Complete set of physical variables (e.g., the arguments of a
wavefunction or quantum state) and dynamical laws (e.g., the
Schr\"{o}dinger equation, or the action for a path integral)
\item The correct solution of the dynamical laws (e.g., the
wavefunction or quantum state of the universe)
\item The connection with observation or experience (e.g., the laws of
psycho-physical parallelism)
\end{enumerate}

Item 1 alone is called a TOE or `theory of everything,' but it is not
complete.  Even 1 and 2 alone are not complete, since they by
themselves do not logically determine what, if any, conscious
experiences occur in a universe.

One can do a Bayesian analysis for the epistemic probabilities of
theories, which include the following elements:

\begin{enumerate}
\item Prior probabilities $p_i\equiv p(T_i)$ for theories $T_i$
\begin{itemize}
\item Necessarily subjective (in my view)
\item Perhaps favoring simplicity, e.g., $p_i = 2^{-n(T_i)}$, where
$n(T_i)$ is an ordering of the theories in increasing order of
complexity
\item Simplicity itself seems subjective
\end{itemize}

\item Conditional probabilities $L_{ij} \equiv P(O_j|T_i)$ of
observations $O_j$ given theories $T_i$ (the likelihoods of the
theories given the observations)

\item Posterior probabilities
\begin{equation}
P_{ij}\equiv P(T_i|O_j)=\frac{P(T_i\& O_j)}{P(O_j)}
=\frac{p(T_i)P(O_j|T_i)}{\sum_k p(T_k)P(O_j|T_k)}
=\frac{p_iL_{ij}}{\sum_k p_kL_{kj}}
\label{Bayes}
\end{equation}
\end{enumerate}

In theories giving many observations, it is controversial what the
conditional probabilities $L_{ij} \equiv P(O_j|T_i)$ should be taken
to be \cite{HS,Pagetyp}.  Is $L_{ij}$ the probability that $O_j$
occurs somewhere in theory $T_i$ \cite{HS}?  In classical mechanics,
this would be $L_{ij}=1$ if $O_j$ occurs and $L_{ij}=0$ if not.  In
conventional quantum mechanics, this would be
$L_{ij}=\langle\psi_i|{\bf{P}}_{O_j}|\psi_i\rangle$, where
$|\psi_i\rangle$ is the quantum state given by the theory $T_1$ and
${\bf{P}}_{O_j}$ is the projection operator onto the observation $O_j$.

This proposal for $L_{ij}$ would not give observational distinctions
between different theories all giving $L_{ij}=1$.  If $T_i$ gives a
universe large enough, many $O_j$'s would give $L_{ij}\approx 1$. 
Then $P_{ij}\equiv P(T_i|O_j)$ would be highest for theories with the
highest $p_i\equiv p(T_i)$, e.g., for the simple theory that all
$O_j$'s certainly exist, with essentially no influence from
observations.

But if some observations occur more in some theory, surely they should
be assigned higher conditional probabilities in that theory.  E.g.,
suppose $T_1$ predicts 1 observation of $O_1$ and $10^6$ observations
of $O_2$, whereas $T_1$ predicts $10^6$ observations of $O_1$ and 1
observation of $O_2$.  Then surely $L_{11} < L_{21}$.  To accomplish
this \cite{Pagetyp}, one might restrict to theories $T_i$ giving
\begin{equation}
\sum_j L_{ij} \equiv \sum_j P(O_j|T_i) = 1,
\label{likenorm}
\end{equation}
so the total probability of all observations is 1 for each theory
$T_1$.

One might expand this rule to a set of principles \cite{Pagetyp} for
Bayesian meta-theories:

\begin{enumerate}
\item {\it Prior Alternatives Principle}:  The set of alternatives to
be assigned conditional probabilities by theories should be chosen
logically prior to the observation to be used to test the theories.

\item {\it Normalization Principle}:  The sum of the conditional
probabilities each theory assigns to all of the alternatives in the
chosen set should be unity.

\item {\it Conscious Anthropic Principle}:  The alternatives ideally
should be chosen to be conscious perceptions or observer-moments.
\end{enumerate}

\section{Sensible Quantum Mechanics}

Conventional quantum mechanics does not seem to obey these principles
\cite{HS}, but {\it Sensible Quantum Mechanics} or {\it Mindless
Sensationalism} \cite{SQM1,SQM2,MS} does.  In it, there is a {\it
quantum world} (what one might call the physical world, excluding
consciousness) with quantum amplitudes and expectation values, but no
probabilities, and there is a {\it conscious world} (what one might
call the mental world) with frequency-type measures or statistical
probabilities.  Interpreting the unconscious quantum world itself
probabilistically is taken to be {\it probabilism}, an {\it
aesthemamorphic myth}.

The axioms of Sensible Quantum Mechanics are the following:

\begin{enumerate}
\item {\it Quantum World Axiom}:  The unconscious ``quantum world''
$Q$ is completely described by an appropriate algebra of operators and
by a suitable state $\sigma$ (a positive linear functional of the
operators) giving the expectation value $\langle A\rangle \equiv
\sigma[A]$ of each operator $A$.

\item {\it Conscious World Axiom}:  The ``conscious world'' $M$, the
set of all conscious perceptions, has a fundamental measure $\mu(S)$
for each subset $S$ of $M$.

\item {\it Quantum-Consciousness Connection}:  The measure $\mu(S)$
for each set $S$ of conscious perceptions is given by the expectation
value of a corresponding ``awareness operator'' $B(S)$, a
positive-operator-valued (POV) measure, in the state $\sigma$ of the
quantum world:
\begin{equation}
\mu(S)=\langle B(S)\rangle \equiv \sigma[B(S)].
\label{CCC}
\end{equation}
\end{enumerate}

If for simplicity we suppose that there is a discrete set of quantum
operators and a discrete set of conscious perceptions or observations
$O_j$, then each Sensible Quantum Mechanics (SQM) theory $T_i$ has

\begin{enumerate}
\item Set of operators $A_j^{(i)}$ and algebra
$A_j^{(i)}A_k^{(i)}=\sum_l c_{jk}^{(i)\ l}A_l^{(i)}$
\item Quantum state $\sigma_1$ giving $\langle A_j^{(i)}\rangle =
\sigma_i[A_j^{(i)}]$.
\item Normalized positive operators $B_j^{(i)}$ for observations
$O_j$, $\sum_j\sigma_i[B_j^{(i)}]=1$.
\item $L_{ij} \equiv P(O_j|T_i) = \sigma_i[B_j^{(i)}] =$ normalized
measure of $O_j$ in $T_i$.
\end{enumerate}

Then if one assigns epistemic prior probabilities $p_i$ for $T_i$ and
has the observation $O_j$, the epistemic posterior probability for
$T_i$ is, by Bayes' theorem,
\begin{equation}
P_{ij} \equiv P(T_i|O_j) 
= \frac{p_i\sigma_i[B_j^{(i)}]}{\sum_k p_k\sigma_k[B_j^{(k)}]}.
\label{postprob}
\end{equation}

The principles of Bayesian meta-theories proposed here and exemplified
by Sensible Quantum Mechanics may be related to various other
principles that have been advocated:

\begin{enumerate}
\item {\it Copernican Principle}:  We are not specially privileged.
\item {\it Weak Anthropic Principle}:  What we observe is conditioned
upon our existence as observers.
\item {\it Principle of Mediocrity}:  We are a ``typical''
civilization \cite{Vil}.
\item {\it Strong Self-Sampling Assumption (SSSA)}:  ``One should
reason as if one's present observer-moment were a random sample from
the set of all observer-moments in its reference class'' \cite{Bos}.
\item {\it Conditional Aesthemic Principle (CAP)}:  ``Unless one has
compelling contrary evidence, one should reason as if one's conscious
perception were a random sample from the set of all conscious
perceptions'' \cite{Pagetyp,SQM1,SQM2,MS}.
\end{enumerate}

\section{Application to Single-History versus Many-Worlds Quantum
Theory}

One can apply these Bayesian meta-theory principles in general, and
the framework of Sensible Quantum Mechanics in particular, to various
issues.  For example, in principle (if the quantum state of the
universe is suitable) one can make statistical tests to distinguish
between single-universe and many-worlds versions of quantum theory
\cite{Page99,Page00,Econ}.

Consider a set of quantum theories giving 99.9\% probability for a
small universe with $N_1=1$ civilization seeing $O_1$, say with
negative cosmological constant $\Lambda < 0$, and 0.1\% probability
for a large universe, $N_2 = 10^6$ civilizations seeing $O_2$, say
with positive cosmological constant $\Lambda > 0$.  Let $T_1$ be a
single-history theory with just a small universe, $T_2$ be a
single-history theory with just a large universe, and $T_3$ a
many-worlds theory with both a small and a large universe.  Say the
respective prior probabilities for these three theories are $p_1$,
$p_2$, and $p_3$, with $p_1+p_2=0.9$ and $p_3=0.1$, so that initially
one gives only 10\% probability for many-worlds.  Since the small
universe is supposed to have 999 times the probability of a large
universe, for the single-universe theories say $p_1 = 0.999(p_1+p_2) =
0.8991$ and $p_2 = 0.001(p_1+p_2) = 0.009$.

In $T_1$, $O_1$ is certain and $O_2$ is impossible, so
\begin{equation}
L_{11} \equiv P(O_1|T_1) = 1,\ L_{12} \equiv P(O_2|T_1) = 0.
\label{T1}
\end{equation}
Conversely, in $T_2$, $O_1$ is impossible and $O_2$ is certain, so
\begin{equation}
L_{21} \equiv P(O_1|T_2) = 0,\ L_{22} \equiv P(O_2|T_2) = 1.
\label{T2}
\end{equation}
On the other hand, in $T_3$, observations are made of both a small
universe and a large universe, so if we take the probabilities of the
observations to be weighted by the number of civilizations making
them, we get
\begin{eqnarray}
L_{31} &\equiv& P(O_1|T_3) = \frac{N_1(0.999)}{N_1(0.999)+N_2(0.001)}
= \frac{0.999}{1000.999} \approx 0.001, \\ \nonumber
L_{32} &\equiv& P(O_2|T_3) = \frac{N_1(0.001)}{N_1(0.999)+N_2(0.001)}
= \frac{1000}{1000.999} \approx 0.999.
\label{T3}
\end{eqnarray}

Suppose then we see $O_2$ (large universe, $\Lambda > 0$).  Then
applying Bayes' theorem gives the posterior probabilities of the three
theories as
\begin{eqnarray}
P(T_1|O_2) &=& \frac{p_1L_{12}}{p_1L_{12}+p_2L_{22}+p_3L_{32}} = 0,
\\ \nonumber
P(T_2|O_2) &=& \frac{p_2L_{22}}{p_1L_{12}+p_2L_{22}+p_3L_{32}} \approx
0.009,
\\ \nonumber
P(T_3|O_2) &=& \frac{p_3L_{32}}{p_1L_{12}+p_2L_{22}+p_3L_{32}} \approx
0.991.
\label{post}
\end{eqnarray}

Thus, in this hypothetical case, even though initially one gave the
many-worlds theory only a 10\% prior probability of being true, after
the observation one would then give it more than 99\% posterior
probability of being true.  Therefore, in principle one can gain
observational evidence of whether or not the many-worlds version of
quantum theory is correct; it is not just an equivalent interpretation
of a single quantum theory.

\section{Application to Fake Universes}

Another application is to fake universes.  Theories with posthuman
civilizations seem in danger of producing too many fake universes
\cite{Bos03}.  Taking substrate-independence of consciousness as given,
Nick Bostrom argues, ``\ldots {\it at least one} of the following
propositions is true:
\begin{enumerate}
\item the humans species is very likely to go extinct before reaching a
`posthuman' stage;
\item any posthuman civilization is extremely unlikely to run a
significant number of simulations of their evolutionary history (or
variations thereof);
\item we are almost certainly living in a computer simulation.''
\end{enumerate}

Paul Davies \cite{Dav} draws the following conclusion from this
argument:  ``For every original world, there will be a stupendous
number of available virtual worlds---some of which would even include
machines simulating virtual worlds of their own, and so on ad
infinitum.''  John Barrow \cite{Bar} makes the further point, ``So we
suggest that, if we live in a simulated reality, we should expect
occasional sudden glitches, small drifts in the supposed constants and
laws of nature over time, and a dawning realization that the flaws of
nature are as important as the laws of nature for our understanding of
true reality.''  If much of posthuman computer simulation is done by
the analogue of today's teenagers, I myself would expect even more
chaos in simulations.

Since we have not observed these postulated sudden glitches, drifts,
and chaos, we might seek explanations of why they are not seen,
incantations to ward off fake universes.  Bostrom himself \cite{Bos03}
noted that humans may go extinct before becoming posthuman, or that
posthumans may be unlikely to run simulations.  But Davies \cite{Dav}
is arguing that these explanations would not ward off fake universes
if there is a multiverse, which he takes as a {\it reductio ad
absurdum} suggesting no multiverse.

In my mind, the weakest assumption leading to fake universes is the
assumption \cite{Bos03} of the substrate-independence of
consciousness. If that is not correct, then it could well be that most
simulated beings simply are not conscious (and so, as conscious
beings, we are not likely to be simulated).  One might analyze this
question within the framework of Sensible Quantum Mechanics, which
postulates that there is a precise positive operator $B_j$ for each
observation or conscious perception $O_j$.  Substrate-independence
seems vague to me but might be taken to mean that $B_j$ is invariant
under some set of transformations (perhaps unitary), such as the
assumption that $B_j = \hat{B}_j = UB_j\tilde{U}$ for some set of
$(U,\tilde{U})$ pairs.  But why should this be true?  It seems highly
implausible to me.

I am reminded of the South Pacific cargo cults, which arose out of
observations during World War II that after airfields and conning
towers were built on their islands, aircraft landed with cargo. 
Thinking that the airfields and conning towers were the sufficient
conditions for such cargo to arrive, they themselves built other
airfields and conning towers, but the expected cargo did not arrive. 
It seems to me that we are at a similar primitive stage concerning what
conditions are sufficient for consciousness (e.g., what the operators
$B_j$ are in Sensible Quantum Mechanics), so guesses that they are
substrate independent are very likely to be quite wrong.

\section{Application to Boltzmann Brains}

Another application is to Boltzmann brains.  Theories in which
spacetime can last too long seem in danger of producing too many
Boltzmann brains (BBs) with mostly disordered observations, which
would make our ordered observations highly atypical
\cite{BB1,BB2,BB3}.

The germ of the idea goes back to Boltzmann \cite{Bolt}, who said that
his ``old assistant'' Dr. Schuetz had suggested that our observed
universe might be a giant thermal fluctuation.  Although this would be
extremely improbable for any particular large region of the universe,
it would certainly occur somewhere if the universe were large enough.
 
However, more recently it has been noticed \cite{Rees,AS} that it
would be much less improbable for a particular region of the universe
to have just a small fluctuation give our observations, rather than
the entire observable region of the universe.  For example, one might
postulate that a human brain arose from a thermal fluctuation in a
state in which it had the same conscious awareness {\it as if} it had
been making observations of a large region of the universe, even if
the large region were {\it actually} not at all what the conscious
awareness thought it was.  (E.g., the conscious perception of the
brain might be that observations had been made of distant stars and
galaxies, whereas actually the surrounding universe might be empty,
without any stars and galaxies, but just having the brain in the same
state it would have been if there had been stars and galaxies it had
observed.)

For a brain of the rough mass of a human brain of one kilogram to
arise from a thermal fluctuation in empty deSitter space at its
Gibbons-Hawking \cite{GH} temperature $T_{\mathrm{dS}}$, the expected
number per 4-volume (3-volume of space multiplied by the time interval
in spacetime) for such long-lasting brains or `long brains' (lbs)
would be
\begin{equation}
\mathcal{N}_\mathrm{lb} \sim
e^{-(1\;\mathrm{kg})c^2/T_\mathrm{dS}} \sim e^{-10^{69}}.
\label{Nlb}
\end{equation}
This is very tiny but would apparently be important if the universe
lasted long enough to make the total 4-volume large in comparison with
$e^{10^{69}}$ times the 4-volume of the region of ordinary observers
(OOs).

I realized \cite{BB1,BB2,BB3} that the rate would be relatively much
larger if the brains were not required to come into long-lasting
existence by a thermal fluctuation but were just required to exist
momentarily as vacuum fluctuations (`brief brains' or bbs).  Then if
the brain were in the right state or configuration, it could have a
brief conscious perception and then disappear into the vacuum again. 
The minimum requirement presumably would be roughly that the matter of
the brain be separated from the corresponding amount of antimatter
that would appear during the virtual loops of the Feynman diagrams of
the vacuum fluctuation.

For a one-kilogram human brain to become separated by its size of say
30 centimeters from the antimatter would require an action of
$(1\;\mathrm{kg})c(0.3\;\mathrm{m})/\hbar \sim 10^{42}$, so the
expected number of brief brains per 4-volume would be of the rough
order of
\begin{equation}
\mathcal{N}_\mathrm{bb} \sim
e^{-(1\;\mathrm{kg})c(0.3\;\mathrm{m})/\hbar} \sim e^{-10^{42}}.
\label{Nbb}
\end{equation}

Although this again is extremely tiny, so that the expected number of
these vacuum fluctuation Boltzmann brains or brief brains (bb BBs)
would be much less than unity over the whole past history of the part
of the universe we can now observe (the `observable universe'), it
would dominate over ordinary observers if the universe lasted long
enough to make the total 4-volume much larger than $e^{10^{42}}$ times
the 4-volume of the region of ordinary observers, which would
apparently be the case if the universe expanded forever.

If some theory predicted that Boltzmann brains greatly dominated over
ordinary observers, we would expect that in that theory we would most
likely be BBs rather than OOs.  Only a very tiny fraction of BBs would
give consciousness (though say still much larger than the number of
OOs), but since we must be conscious to observe anything, this
condition is a necessary selection effect, so that if such conscious
BBs and/or OOs exist, it would not be at all improbable that we are a
conscious BB or OO (and most likely a BB if there are far more
conscious experiences produced by them, say in a universe that lasts
much longer and grows much bigger than the region where OOs can
exist).

However, once we include the necessary selection effect that we are
conscious, the conditional probabilities for the various possible
natures of the conscious perceptions should be given by the theory. 
If BBs enormously dominate over OOs, we should expect that we are BBs
rather than OOs (or rather you should, since then most probably I am
just a figment of your imagination).  But only a very tiny fraction of
BB conscious perceptions would be expected to be so ordered as yours
generally are, with coherent visual images and memories, so the fact
that yours are highly ordered counts as strong statistical evidence
against any theory that implies that almost all conscious perceptions
are produced by BBs (since they would produce mostly disordered
observations).

If one includes more details of the order that you observe, then the
likelihood that your observation comes from a theory in which BBs
dominate is even lower.  For example, even among the tiny fraction of
BB conscious perceptions that are ordered, only a tiny fraction would
give ordered observations of, say, galaxy-galaxy correlations.  And of
these, only a tiny fraction would give a galaxy-galaxy correlation
similar to what you observe or may be aware of from what you think are
the observations of others.  So if you put in all the detail of the
order that you observe or are aware of, it would be very unlikely that
it is a conscious perception produced in a theory in which Boltzmann
brains greatly dominate.

Although the next step would be going beyond what could be
observationally checked, one might note that even among the extremely
tiny fraction of BB conscious perceptions that included all the
details of the order that you perceive, only a very tiny fraction of
these would be non-illusory, say of actual galaxies, since there would
be far more BBs in the entire huge spacetime to have brains in just
the right states to have illusory perceptions that they have observed
galaxies with all the correlations that you observe or are aware of,
than for all the `observed' galaxies actually to exist.

In conclusion, theories in which Boltzmann brains greatly dominate
over ordinary observers would make your or my observations highly
atypical.  Some might argue \cite{HS} that this atypicality does not
imply a correspondingly small likelihood for these theories, but I
would respond \cite{Pagetyp} that it should.

Let me give some numerical estimates for the conditions that Boltzmann
brains might dominate ordinary observers in our universe.  Basically,
a sufficient condition would seem to be that the expectation value of
the total 4-volume of our comoving region of the universe is infinite
when one includes the future.

After the period of radiation domination ended when the linear size of
a comoving region of our universe was thousands of times smaller than
today, our observed comoving region (that given by fixed markers such
as galaxies that move apart with the expansion of the universe) is
apparently dominated by a cosmological constant $\Lambda$ and by
low-velocity massive particles with negligible pressure (called `dust'
in cosmology), a cold-dark-matter-$\Lambda$ model or CDM$\Lambda$
model.  Its spacetime geometry is apparently well described by the
$k=0$ (spatially flat) Friedmann-Robertson-Walker metric with
unit-jerk ($J\equiv a^2(d^3a/dt^3)/(da/dt)^3 = 1$, where $a(t)$ is the
linear size of a fixed comoving region as a function of the time $t$),
\begin{equation}
ds^2 = T^2[-d\tau^2 + (\sinh^{4/3}{\tau}) (dr^4 + r^2 d\theta^2 +
r^2\sin^2{\theta} d\varphi^2)].
\label{metric}
\end{equation}

The Hubble Space Telescope key project \cite{key} gives that the
present Hubble constant of our universe, today's value of $H \equiv
d\ln{a}/dt$, is $H_0 = 72\pm 8$ km/s/Mpc.  From the third-year results
of the Wilkinson Microwave Anisotropy Probe (WMAP), the fraction of
the critical energy density given by the cosmological constant is
$\Omega_\Lambda = 0.72\pm0.04$.  Putting these numbers together
\cite{BB1,BB3} gives that the asymptotic value of the Hubble
expansion rate in the distant future is
\begin{equation}
H_\Lambda = \sqrt{\Lambda/3} = H_0\sqrt{\Omega_\Lambda} \approx (16\pm
2\ \mathrm{Gyr})^{-1},
\label{Hlambda}
\end{equation}
the value of the constant parameter $T$ in the metric above is
\begin{equation}
T = (2/3)H_\Lambda^{-1} = 11\pm 1.5\ \mathrm{Gyr},
\label{T}
\end{equation}
the present value of the dimensionless time coordinate $\tau$ in the
metric above is
\begin{equation}
\tau_0 = \tanh^{-1}{\sqrt{\Omega_\Lambda}} \approx 1.25\pm 0.09,
\label{tau}
\end{equation}
and the present value of the Hubble constant, $H_0$, multiplied by the
present value of the age, $t_0$, or the present value of
$d\ln{a}/d\ln{t}$, is
\begin{equation}
H_0 t_0 = \frac{2}{3}
\frac{\tanh^{-1}{\sqrt{\Omega_\Lambda}}}{\sqrt{\Omega_\Lambda}}
\approx 0.99\pm 0.04.
\label{Ht}
\end{equation}

It is interesting that this last value is consistent with unity, up to
a fraction of the current observational uncertainties.  This means
that when one plots the linear size $a$ versus the time $t$ since the
beginning in this model (applicable long after very early inflation,
and also after radiation domination ended), the $a(t)$ curve bends
downward during the first part of the history (deceleration, when dark
matter dominated and made gravity attractive) and then bends upward
during the latter part of the past history (acceleration, when the
cosmological constant or dark energy dominated and made gravity
repulsive on cosmological scales) in just such a way that if one draws
a straight line tangent to the $a(t)$ curve at the present time $t_0$,
it will have slope $da/dt$ that is very nearly the same as $a/t$ and
hence will very nearly go through the origin.  To put it another way,
not counting the very early inflationary period, the value of linear
size over age, $a/t$, is, within the current observational
uncertainties, minimized at the present time.  This coincidence, along
with the observed spatial flatness, plus the fact that the current
best estimate \cite{WMAP} of the age 13.7 Gyr (Gigayears or billions
of years) has the same first three digits as the reciprocal of the
fine structure constant, $1/\alpha \approx 137.036$, allows one to
write the metric and parameters above in an easy-to-remember form that
I might call the {\it mnemonic universe}:  CDM$\Lambda$ $k=0$ FRW $t_0
= H_0^{-1} = 10^8\ \mathrm{yr}/\alpha$.

If this expanding universe lasts forever, then for any fixed comoving
volume (e.g., the physically expanding volume occupied by some fixed
set of galaxies and whatever massive particles they may eventually
decay into, say electrons and positrons) that presumably contains only
a finite number of ordinary observers (say if they cannot exist after
the stars burn out in a finite time), one will have an infinite amount
of 4-volume in the infinite future, and hence apparently an infinite
number of Boltzmann brains, infinitely dominating over the presumably
finite number of ordinary observers.  One might however postulate that
the universe will not expand forever but perhaps decay to some
entirely different configuration obeying quite different effective
laws of physics (say not giving any conscious brains at all) by the
quantum formation of a bubble that then expands at the speed of light
and destroys all that it engulfs.  If the bubble nucleation rate per
4-volume is $A$ (for `annihilation,' which is what the bubble would do
to the present form and laws of physics of the universe), then the
expectation value of the total 4-volume per fixed comoving 3-volume
would be finite only if the bubble formation rate is greater than some
calculable minimum \cite{BB1,BB3}:
\begin{equation}
A > A_\mathrm{min} = \frac{9}{4\pi} H_\Lambda^4 =
\frac{\Lambda^4}{4\pi} \approx (18\pm 2\ \mathrm{Gyr})^{-4} \sim
e^{-563},
\label{Amin}
\end{equation}
where the last number is in Planck units, setting $\hbar = c = G =
1$.

For a given value of $A$, the survival probability of the CDM$\Lambda$
$k=0$ FRW  model above is
\begin{equation}
P(t) = \exp{\left[-\frac{16}{27}\frac{A}{A_{\mathrm min}}
        \int_0^{\frac{3}{2}H_\Lambda t} dx \sinh^2{x}
         \left(\int_x^{\frac{3}{2}H_\Lambda t}
	  \frac{dy}{\sinh^{2/3}{y}}\right)^3\right]}.
\label{survival}
\end{equation}
Given that we have lasted until today, the probability per year for
being annihilated by a bubble that has formed just outside our past
light cone (and hence which will engulf us within the coming year,
though we cannot see any signal of its coming since it would be coming
at essentially the speed of light) would be greater than one part in
one hundred billion.  With the present earth population of nearly 7
billion, this would give an minimal expected death rate of about 7
persons per century.  (Of course, it could not be 7 persons in one
century, but all 7 billion with a probability of about one in a
billion per century.)  It also gives an upper limit on the present
half-life of our universe of 19 Gyr \cite{BB1}.

Since we do not see that our observations are so disordered as would be
statistically predicted if Boltzmann brains greatly dominated over
ordinary observers, we might seek an explanation of why they apparently
do not dominate, incantations to ward off Boltzmann brains:
\begin{enumerate}
\item The universe may be decaying classically if the quintessence
potential slowly slides negative, so that the universe rolls into
oblivion \cite{Page06}.
\item The universe may be decaying quantum mechanically with $e^{-563}
\stackrel{<}{\sim} A \stackrel{<}{\sim} e^{-556}$, though this seems to
require unnatural fine tuning \cite{BB1,BB2,BB3}.
\item Observers might conceivably fill the entire universe and be too
large to form by vacuum fluctuations \cite{BB1,BB2,BB3}.
\item The local view restricts the 3-volume to the causal patch
\cite{BF}.
\item Use a regularization effectively cutting off before the Boltzmann
brains \cite{Linde}
\item Boltzmann brains should be lumped with bubble formation
\cite{Vilenkin}.
\item Holographic cosmology gives a time-dependent Hamiltonian
\cite{Banks}.
\item `Constants' of physics change so that the rate per 4-volume of
Boltzmann brains asymptotically decays to zero \cite{Carlip}.
\item Restrict to only certain group-averaged quantum states
\cite{GM}.
\item Quantum field theory may not apply to brains \cite{BB1,BB3}.
\item We may not be typical \cite{HS}.
\end{enumerate}

Since an astronomical decay rate is suggested by one way of
regularizing Boltzmann brains, and since it is not completely ruled
out, perhaps we should take it seriously as one possibility (though not
the only one).  So let us see what are implications of the possible
astronomical decay rate.

If ordinary observers could last until engulfed by a bubble:
\begin{enumerate}
\item They could not see it coming and so would not dread it.
\item They would be destroyed so fast they would feel no pain.
\item There would be no grieving survivors left behind.
\end{enumerate}
So it would be the most humanely possible execution.

Furthermore, the universe will always persist in some decreasing
fraction or measure of Everett worlds.  Thus one could never absolutely
rule out a decaying universe by observations at any finite time, though
at sufficiently late times observations would be strong statistical
evidence against the astronomical decay rate.

The main point of my discussing this scenario is not to advocate a
particular solution to the Boltzmann brain problem, but rather to
stimulate more research on the huge scientific mystery of the measure
for the string/M landscape \cite{Sussbook,Vilbook} or other multiverse
theory \cite{multiverse}.

\section{Application to Biological Evolution}

Sensible Quantum Mechanics suggests we are effectively selected
randomly by our measure of consciousness, like winners in a cosmic
jackpot \cite{Davies}.  No further selection would be needed {\it
within} a particular SQM theory.

This does not explain {\it which} SQM theory is correct.  One might
suppose it was chosen by a highly lawful and yet benevolent God
(perhaps ultimately simple \cite{Swinburne}, say the greatest possible
being, {\it contra} Dawkins' idea that God is complex \cite{Dawkins}),
who wanted to create understanding conscious beings in His image
within a highly ordered and elegant multiverse.

Within Darwinian evolution on earth, our ancestors would not just be
random life but conditionalized by being our ancestors.  One or more
correlated steps might be highly improbable on a per-planet basis
\cite{Carter} (say $P \sim 10^{-n}$ with $n \gg 24$ so that the
probability would be very low for any others within the observable
universe of probably not more than $10^{24}$ planets).  Only steps off
our ancestral line can be argued to be probable to occur on a random
planet in which the steps up to that point have occurred.  It would be
interesting to see what features of complexity and intelligence
developed off our ancestral line (e.g., to octopi from our last common
ancestor), and what features have developed only within our ancestral
line.  Some of the latter features could be so rare that they have not
occurred anywhere else within the observable universe (the part we can
see), even though the entire universe may be so vast that all of these
features have occurred many times very much further away than we can
possibly see.

\section{Conclusions}

\begin{enumerate}
\item Scientific theories need to be testable by observations.
\item Multiverse theories can in principle be testable if they give
probabilities for observations that depend upon the theory.
\item Sensible Quantum Mechanics (SQM) gives probabilities that are
expectation values of positive operators associated with observations
or conscious perceptions, and in principle they can be tested
observationally.
\item One can test between single-history and many-worlds theories.
\item Fake universe and Boltzmann brains are threats to certain
multiverse (and single-universe) theories.
\item Sensible Quantum Mechanics suggests that fake universes may not
be a threat, since there is no compelling reason to believe in the
substrate independence of consciousness.
\item There are many suggested solutions for Boltzmann brains, but it
just is not clear what the correct one is.
\item No further life principle is needed {\it within} a Sensible
Quantum Mechanics theory, but {\it which} Sensible Quantum Mechanics
theory is correct needs explaining.
\item From the view of foresight, our biological ancestors may be
highly improbable, post-selected by our existence as beings with a high
measure of conscious perceptions.
\end{enumerate}

\section*{Acknowledgments}

\hspace{.20in} I am indebted to discussions with Andreas Albrecht,
Denis Alexander, John Barrow, Nick Bostrom, Raphael Bousso, Andrew
Briggs, Peter Bussey, Bernard Carr, Sean Carroll, Brandon Carter,
Kelly James Clark, Gerald Cleaver, Francis Collins, Robin Collins,
Gary Colwell, William Lane Craig, Paul Davies, Richard Dawkins,
William Dembski, David Deutsch, Michael Douglas, George Ellis, Debra
Fisher, Charles Don Geilker, Gary Gibbons, J.~Richard Gott, Thomas
Greenlee, Alan Guth, James Hartle, Stephen Hawking, Rodney Holder,
Richard Hudson, Chris Isham, Renata Kallosh, Denis Lamoureux, John
Leslie, Andrei Linde, Robert Mann, Don Marolf, Greg Martin, Alister
McGrath, Gerard Nienhuis, Gary Patterson, Alvin Plantinga, Chris
Polachic, John Polkinghorne, Martin Rees, Hugh Ross, Peter Sarnak,
Henry F.~Schaefer III, James Sinclair, Lee Smolin, Mark Srednicki, Mel
Stewart, Jonathan Strand, Leonard Susskind, Richard Swinburne, Max
Tegmark, Donald Turner, Neil Turok, Bill Unruh, Alex Vilenkin, Steven
Weinberg, and others whom I don't recall right now, on various aspects
of this general issue, though the opinions expressed are my own.  My
scientific research on the quantum cosmology and the multiverse is
supported in part by the Natural Sciences and Research Council of
Canada.


\newpage
\baselineskip 5pt

\begin{thebibliography}{99}


\bibitem{HS} James B.~Hartle and Mark Srednicki, ``Are We Typical?''
{\it Physical Review} {\bf D75}: 123523 (2007) [eprint arXiv:0704.2630
$<$http://arxiv.org/abs/0704.2630$>$].

\bibitem{Pagetyp} Don N.~Page, ``Typicality Defended,'' [eprint
arXiv:0707.4169 $<$http://arxiv.org/abs/0707.4169$>$].

\bibitem{SQM1} Don N.~Page, ``Sensible Quantum Mechanics: Are Only
Perceptions Probabilistic?'' [eprint arXiv:quant-ph/9506010
$<$http://arxiv.org/abs/quant-ph/9506010$>$].

\bibitem{SQM2} Don N.~Page, ``Sensible Quantum Mechanics: Are
Probabilities Only in the Mind?'' {\it International Journal of Modern
Physics} {\bf D5}, 583-596 (1996) [eprint arXiv:gr-qc/9507024
$<$http://arxiv.org/abs/gr-qc/9507024$>$].

\bibitem{MS} Don N.~Page, ``Mindless Sensationalism: A Quantum
Framework for Consciousness,''in {\em Consciousness: New Philosophical
Perspectives}, edited by Quentin Smith and Alexsandar Jokic (Oxford,
Oxford University Press, 2003), pp.\ 468-506 [eprint
arXiv:quant-ph/0108039 $<$http://arxiv.org/abs/quant-ph/0108039$>$].

\bibitem{Vil} Alexander Vilenkin, ``Predictions from Quantum
Cosmology,'' {\it Physical Review Letters} {\bf 74}, 846-849 (1995)
[eprint arXiv:gr-qc/9406010 $<$http://arxiv.org/abs/gr-qc/9406010$>$].

\bibitem{Bos} Nick Bostrom, {\em Anthropic Bias:  Observation
Selection Effects in Science and Philosophy} (Routledge, New York,
2002).

\bibitem{Page99} Don N.~Page, ``Observational Consequences of
Many-World Quantum Theory,'' [eprint arXiv:quant-ph/9904004
$<$http://arxiv.org/abs/quant-ph/9904004$>$].

\bibitem{Page00} Don N.~Page, "Can Quantum Cosmology Give Observational
Consequences of Many-Worlds Quantum Theory?" in {\em General
Relativity and Relativistic Astrophysics, Eighth Canadian Conference,
Montreal, Quebec, 1999}, edited by Cliff P.~Burgess and Robert
C.~Myers (American Institute of Physics, Melville, New York, 1999),
pp.~225-232 [eprint arXiv:gr-qc/0001001
$<$http://arxiv.org/abs/gr-qc/0001001$>$].

\bibitem{Econ} Meher Antia, {\em The Economist}, May 22, 1999, p.~145.

\bibitem{Bos03} Nick Bostrom, ``Are You Living in a Computer
Simulation?'' {\it Philosophical Quarterly} {\bf 53}, 243-255 (2003);
$<$www.simulation-argument.com$>$.

\bibitem{Dav} Paul Davies, ``A Brief History of the Multiverse,'' {\it
New York Times}, April 12, 2003.

\bibitem{Bar} John D.~Barrow, ``Living in a Simulated Universe,'' in
{\em Universe or Multiverse?}, edited by Bernard Carr (Cambridge
University Press, Cambridge, 2007), pp.~481-486.

\bibitem{BB1} Don N.~Page, ``Is Our Universe Likely to Decay within 20
Billion Years?'' [eprint arXiv:hep-th/0610079
$<$http://arxiv.org/abs/hep-th/0610079$>$].

\bibitem{BB2} Don N.~Page, ``Return of the Boltzmann Brains,'' [eprint
arXiv:hep-th/0611158 $<$http://arxiv.org/abs/hep-th/0611158$>$].

\bibitem{BB3} Don N.~Page, ``Is Our Universe Decaying at an
Astronomical Rate?'' [eprint arXiv:hep-th/0612137
$<$http://arxiv.org/abs/hep-th/0612137$>$].

\bibitem{Bolt} Ludwig Boltzmann, {\it Nature} {\bf 51}, 413-415 (1895).

\bibitem{Rees} Martin J.~Rees, {\em Before the Beginning:  Our
Universe and Others} (Simon and Schuster, New York, 1997), p.~221.

\bibitem{AS} Andreas Albrecht and Lorenzo Sorbo, {\it Physical Review}
{\bf D70}: 063528 (2004); eprint hep-th/0405270
$<$http://arxiv.org/abs/hep-th/0405270$>$.

\bibitem{GH} Gary W.~Gibbons and Stephen W.~Hawking,
{\it Physical Review} {\bf D15}, 2738-2751 (1977).

\bibitem{key} Wendy L.~Freedman, B.~F.~Madore, B.~K.~Gibson,
L.~Ferrarese, D.~D.~Kelson, S.~Sakai, J.~R.~Mould, R.~C.~Kennicutt,
H.~C.~Ford, J.~A.~Graham, J.~P.~Huchra, S.~M.~G.~Hughes, Garth
D.~Illingworth, L.~M.~Macri, and P.~B.~Stetson, ``Final Results from
the Hubble Space Telescope Key Project to Measure the Hubble
Constant,'' {\it Astrophysical Journal} {\bf 553}, 47-72 (2001)
[eprint arXiv:astro-ph/0012376
$<$http://arxiv.org/abs/astro-ph/0012376$>$].

\bibitem{WMAP} David N.~Spergel, Licia Verde, Hiranya V.~Peiris,
E.~Komatsu, M.~R.~Nolta, C.~L.~Bennett, M.~Halpern, G.~Hinshaw,
N.~Jarosik, A.~Kogut, M.~Limon, S.~S.~Meyer, L.~Page, G.~S.~Tucker,
J.~L.~Weiland, E.~Wollack, and E.~L.~Wright, ``First Year Wilkinson
Microwave Anisotropy Probe (WMAP) Observations: Determination of
Cosmological Parameters,'' {\it Astrophysical Journal Supplement
Series} {\bf 148}, 175-194 (2003) [eprint arXiv:astro-ph/0302209
$<$http://arxiv.org/abs/astro-ph/0302209$>$].

\bibitem{Page06} D.~N.~Page, ``The Lifetime of the Universe,'' {\it
Journal of the Korean Physical Society} {\bf 49}, 711-714 (2006)
[eprint arXiv:hep-th/0510003
$<$http://arxiv.org/abs/hep-th/0510003$>$].

\bibitem{BF} Raphael Bousso and Ben Freivogel, ``A Paradox in the
Global Description of the Multiverse,'' {\it Journal of High Energy
Physics} {\bf 0706}: 018 (2007) [eprint arXiv:hep-th/0610132
$<$http://arxiv.org/abs/hep-th/0610132$>$].

\bibitem{Linde} Andrei Linde, ``Sinks in the Landscape, Boltzmann
Brains, and the Cosmological Constant Problem,'' {\it Journal of
Cosmological and Astroparticle Physics} {\bf 0701}: 022 (2007) [eprint
arXiv:hep-th/0611043 $<$http://arxiv.org/abs/hep-th/0611043$>$].

\bibitem{Vilenkin} Alexander Vilenkin, ``Freak Observers and the
Measure of the Multiverse,'' {\it Journal of High Energy Physics} {\bf
0701}: 092 (2007) [eprint arXiv:hep-th/0611271
$<$http://arxiv.org/abs/hep-th/0611271$>$].

\bibitem{Banks} Thomas Banks, ``Entropy and Initial Conditions in
Cosmology,'' [eprint arXiv:hep-th/0701146
$<$http://arxiv.org/abs/hep-th/0701146$>$].

\bibitem{Carlip} Steve Carlip, ``Transient Observers and Variable
Constants, or Repelling the Invasion of the Boltzmann's Brains,'' {\it
Journal of Cosmological and Astroparticle Physics} {\bf 0706}: 001
(2007) [eprint arxiv:hep-th/0703115
$<$http://arxiv.org/abs/hep-th/0703115$>$].

\bibitem{GM} Steven B.~Giddings and Donald Marolf, ``A Global Picture
of Quantum de Sitter Space,'' {\it Physical Review} {\bf D76}: 064023
(2007) [eprint: arXiv:hep-th/0705.1178
$<$http://arxiv.org/abs/0705.1178$>$].

\bibitem{Sussbook} Leonard Susskind, {\em The Cosmic Landscape:  String
Theory and the Illusion of Intelligent Design} (Little, Brown and
Company, New York, 2006).

\bibitem{Vilbook} Alex Vilenkin, {\em Many Worlds in One: The Search
for Other Universes} (Hill and Wang, New York, 2006).

\bibitem{multiverse} Bernard Carr, editor, {\em Universe or
Multiverse?} (Cambridge University Press, Cambridge, 2007).

\bibitem{Davies} Paul Davies, {\em Cosmic Jackpot:  Why Our Universe
Is Just Right for Life} (Houghton Mifflin, Boston, 2007).

\bibitem{Swinburne} Richard Swinburne, {\em The Existence of God}, 2nd
edition (Clarendon Press, Oxford, 2004).

\bibitem{Dawkins} Richard Dawkins, {\em The God Delusion} (Houghton
Mifflin, Boston, 2006).

\bibitem{Carter} Brandon Carter, ``The Anthropic Principle and its
Implications for Biological Evolution,'' {\it Philosophical
Transactions of the Royal Society of London} {\bf A310}, 347-363
(1983).


\end{thebibliography}
\end{document}